\def \VersionAuthor {}
\ifdefined\VersionAuthor
	\newcommand{\AuthorVersion}[1]{#1}
	\newcommand{\FinalVersion}[1]{}
\else
	\newcommand{\AuthorVersion}[1]{}
	\newcommand{\FinalVersion}[1]{#1}
\fi

\documentclass[a4paper,10pt,runningheads]{llncs}
\makeatletter
\AtBeginDocument{%
  \@ifpackageloaded{hyperref}
  {\def\@doi#1{\href{https://doi.org/#1}
      {\ttfamily https://doi.org/#1}\egroup}}
  {\def\@doi#1{\ttfamily https://doi.org/#1\egroup}}
  \def\doi{\bgroup\catcode`\_=12\relax\@doi}}
\makeatother
\makeatletter
\def\@biblabel#1{[#1]}

\makeatother

\usepackage[utf8]{inputenc}
\usepackage[english]{babel} %
\usepackage{csquotes}
\usepackage{lmodern} %
\usepackage[T1]{fontenc} %

\usepackage[ruled,vlined,linesnumbered]{algorithm2e}
	\SetKwInOut{Input}{input}
	\SetKwInOut{Output}{output}

\usepackage{subcaption}

\usepackage{amsmath} %
\usepackage{amssymb} %
\usepackage{mathtools} %

\usepackage[misc,geometry]{ifsym} %

\usepackage{paralist} %

\newenvironment{ienumerate}
	{\ifdefined\VersionLong\begin{enumerate}\else\begin{inparaenum}[\itshape i\upshape)]\fi}
	{\ifdefined\VersionLong\end{enumerate}\else\end{inparaenum}\fi}

\ifdefined \VersionLong
	\newcommand{\LongVersion}[1]{#1}
	\newcommand{\ShortVersion}[1]{}
\else
	\newcommand{\LongVersion}[1]{}
	\newcommand{\ShortVersion}[1]{#1}
\fi

\ifdefined\VersionAuthor
	\usepackage[backend=biber,backref=true,style=alphabetic,url=false,doi=true,defernumbers=true,sorting=anyt,maxnames=99]{biblatex} %
	\DeclareSourcemap{
		\maps[datatype=bibtex]{
			\map[overwrite]{
				\step[fieldsource=toappear, match={true},
					fieldset=note, fieldvalue={To appear.~{}}, append]
			}
		}
	}

	\renewbibmacro*{doi+eprint+url}{%
		\iftoggle{bbx:doi}
			{\color{black!40}\footnotesize\printfield{doi}}
			{}%
		\newunit\newblock
		\iftoggle{bbx:eprint}
			{\usebibmacro{eprint}}
			{}%
		\newunit\newblock
		\iftoggle{bbx:url}
			{\usebibmacro{url+urldate}}
			{}%
	}

\fi
\ifdefined\VersionWithComments
	\usepackage{draftwatermark}
	\SetWatermarkText{draft}
	\SetWatermarkScale{2}
	\SetWatermarkColor[gray]{0.9}
\fi
\usepackage[svgnames,table]{xcolor}
\definecolor{USPNcobalt}{HTML}{293358}
\definecolor{USPNocre}{HTML}{8b7d6d}
\definecolor{USPNblanc}{HTML}{ffffff}
\definecolor{USPNceruleen}{HTML}{354878}
\definecolor{USPNsable}{HTML}{ad947e}

\usepackage[
		pdfauthor={Etienne Andre},%
		pdftitle={Tuning Trains Speed in Railway Scheduling},
		breaklinks  = true,
		colorlinks  = true,
	\ifdefined \VersionWithComments
	\fi
		citecolor   = USPNsable,
		linkcolor   = USPNocre,
		urlcolor    = USPNceruleen,
	]{hyperref}

\usepackage[capitalise,english,nameinlink]{cleveref} %
\crefname{line}{\text{line}}{\text{lines}} %

\newcommand{\defProblem}[3]
{%
\noindent\fcolorbox{black}{USPNsable!15}{
	\begin{minipage}{.95\columnwidth}
		\textbf{#1:}\\ %
		\textsc{Input}: #2\\
		\textsc{Problem}: #3
	\end{minipage}
}

	\smallskip

}

\usepackage{tikz}
\usetikzlibrary{arrows,automata}

\tikzstyle{pta}=[auto, ->, >=stealth']
\tikzstyle{every node}=[initial text=]
\tikzstyle{location}=[rectangle, rounded corners, minimum size=12pt, draw=black, fill=blue!10, inner sep=2pt]
\tikzstyle{invariant}=[draw=black, dotted, inner sep=1pt, node distance=0] %
\tikzstyle{final}=[double, fill=blue!50]
\tikzstyle{edge}=[->]

\definecolor{coloract}{rgb}{0, 0.3, 0}
\definecolor{colorclock}{rgb}{0.7, 0, 0}
\definecolor{colordisc}{rgb}{1, 0, 1}
\definecolor{colorloc}{rgb}{0.4, 0.4, 0.65}
\definecolor{colorparam}{rgb}{1, 0.6, 0.0}

\tikzstyle{railway}=[auto] %
\tikzstyle{boundary}=[circle, minimum size=12pt, draw=black, fill=pink, inner sep=2pt]
\tikzstyle{railwaynode}=[circle, minimum size=0pt, inner sep=-2pt]
\tikzstyle{station}=[circle, inner sep=-2pt]

\newcommand{\pacman}[3]{\tikz[baseline, #1]{%
    \fill[black] (0, 0) circle (1cm);
    \draw[thick,fill=#2]
    (0,0) -- (60:1cm) arc (60:300:1cm) -- cycle;
    \node at (0,0) {\textcolor{white}{\textbf{#3}}};
    }
    }

\newcommand{\railwaynode}[1]{\pacman{#1}{blue!50}{}}
\newcommand{\station}[2]{\pacman{#1}{red!75!black}{#2}}

\usetikzlibrary{trambws}
\newcommand{\trainTikZ}[2]{%
	\begin{tikzpicture}
		\pic [lliw y tram=#2] {tram={#1}};
	\end{tikzpicture}
}

\newcommand{\rowHeader}{\rowcolor{USPNsable!50}}
\newcommand{\cellCenter}[1]{\multicolumn{1}{c|}{#1}}

\newcommand{\cellTO}{\cellcolor{red!40}\textbf{T.O.}}

\newcommand{\assign}{\leftarrow}
\newcommand{\topartial}{\rightharpoonup}

\newcommand{\checkUseMacro}[1]{#1}

\newcommand{\setN}{\ensuremath{\mathbb{N}}}
\newcommand{\setQ}{\ensuremath{\mathbb{Q}}}
\newcommand{\setQgeqzero}{\ensuremath{\setQ_{\geq 0}}}
\newcommand{\setR}{\ensuremath{\mathbb{R}}}
\newcommand{\setRgeqzero}{\ensuremath{\setR_{\geq 0}}}

\newcommand{\setZ}{\ensuremath{\mathbb{Z}}}

\newcommand{\compOp}{\bowtie}
\newcommand{\init}{\ensuremath{0}}
\newcommand{\final}{\ensuremath{f}}

\newcommand{\valuate}[2]{\ensuremath{#2(#1)}}

\newcommand{\BTrue}{\ensuremath{\top}}
\newcommand{\BFalse}{\ensuremath{\bot}}

\newcommand{\styleAutomaton}[1]{\ensuremath{\mathcal{#1}}}
\newcommand{\textstyleact}[1]{\ensuremath{\mathit{#1}}}
\newcommand{\textstyleclock}[1]{\ensuremath{\mathit{#1}}}
\newcommand{\textstyledisc}[1]{\ensuremath{\mathit{#1}}}
\newcommand{\textstyleloc}[1]{\ensuremath{\mathrm{#1}}}
\newcommand{\textstyleparam}[1]{\ensuremath{{#1}}}

\newcommand{\styleact}[1]{\textcolor{coloract}{\textstyleact{#1}}}
\newcommand{\styledisc}[1]{\textcolor{colordisc}{\textstyledisc{#1}}}
\newcommand{\styleclock}[1]{\textcolor{colorclock}{\textstyleclock{#1}}}

\newcommand{\clock}{\ensuremath{\textstyleclock{x}}}

\newcommand{\clocki}[1]{\ensuremath{\textstyleclock{\clock_{#1}}}}
\newcommand{\ClockCard}{H} %
\newcommand{\clockval}{\ensuremath{\mu}}
\newcommand{\ClockSet}{\ensuremath{\mathbb{X}}} %
\newcommand{\ClocksZero}{\ensuremath{\vec{0}}}
\newcommand{\clockabs}{\ensuremath{\textstyleclock{\clock_{\mathit{abs}}}}}

\newcommand{\segfreei}[1]{\ensuremath{\textstyledisc{segfree_{#1}}}}

\newcommand{\resets}{\ensuremath{R}}
\newcommand{\reset}[2]{\ensuremath{[#1]_{#2}}}

\newcommand{\param}{\ensuremath{\textstyleparam{p}}}
\newcommand{\parami}[1]{\ensuremath{\textstyleparam{\param_{#1}}}}
\newcommand{\ParamCard}{\ensuremath{M}} %
\newcommand{\pval}{\ensuremath{v}}
\newcommand{\ParamSet}{\ensuremath{\mathbb{P}}} %

\newcommand{\paramBound}{\ensuremath{\mathit{bnd}}}
\newcommand{\paramJ}{\ensuremath{J}}
\newcommand{\paramRed}{\ensuremath{\parami{R}}}

\newcommand{\TA}{\ensuremath{\checkUseMacro{\styleAutomaton{A}}}}
\newcommand{\PTA}{\ensuremath{\checkUseMacro{\styleAutomaton{A}}}}

\newcommand{\action}{\ensuremath{\textstyleact{a}}}
\newcommand{\ActionSet}{\ensuremath{\Sigma}}
\newcommand{\constraint}{\ensuremath{C}}

\newcommand{\edge}{\ensuremath{\checkUseMacro{e}}}

\newcommand{\EdgeSet}{\ensuremath{E}}

\newcommand{\guard}{\ensuremath{g}}

\newcommand{\invariant}{\ensuremath{I}}

\newcommand{\loc}{\ensuremath{\textstyleloc{\ell}}}
\newcommand{\loci}[1]{\ensuremath{\textstyleloc{\loc_{#1}}}}
\newcommand{\locinit}{\ensuremath{\textstyleloc{\loc_\init}}}
\newcommand{\locfinal}{\ensuremath{\textstyleloc{\loc_\final}}}
\newcommand{\LocSet}{\ensuremath{L}} %
\newcommand{\longuefleche}[1]{\stackrel{#1}{\longrightarrow}}

\newcommand{\semantics}[1]{\ensuremath{\mathfrak{T}_{#1}}}

\newcommand{\transition}{{\ensuremath{\rightarrow}}}
\newcommand{\transitionWith}[1]{\stackrel{#1}{\mapsto}}

\newcommand{\StateSet}{\ensuremath{\mathfrak{S}}}
\newcommand{\concstate}{\ensuremath{\mathfrak{s}}}

\newcommand{\concstateinit}{\ensuremath{\concstate_\init}}

\newcommand{\railGraph}{\ensuremath{\mathcal{G}}}
\newcommand{\railBoundaries}{\ensuremath{B}}
\newcommand{\railDuration}{\ensuremath{\mathit{SegDur}}}
\newcommand{\railDurationPair}{\ensuremath{\mathit{SegsDur}}}
\newcommand{\railNode}{\ensuremath{n}}
\newcommand{\railNodes}{\ensuremath{N}}
\newcommand{\railSegment}{\ensuremath{\mathit{seg}}}
\newcommand{\railSegments}{\ensuremath{\mathit{Seg}}}
\newcommand{\railStations}{\ensuremath{\mathit{St}}}
\newcommand{\railTransitions}{\ensuremath{\mathit{T}}}
\newcommand{\train}{\ensuremath{t}}
\newcommand{\Trains}{\ensuremath{\mathcal{T}}}
\newcommand{\trainConnection}{\ensuremath{\mathit{C}}}
\newcommand{\trainDuration}{\ensuremath{\mathit{TSegDur}}}
\newcommand{\trainDurationPair}{\ensuremath{\mathit{TSegsDur}}}
\newcommand{\trainBlue}{\ensuremath{\train_{\mathit{blue}}}}
\newcommand{\trainGreen}{\ensuremath{\train_{\mathit{green}}}}
\newcommand{\trainRed}{\ensuremath{\train_{\mathit{red}}}}

\newcommand{\ScheduleConstraints}{\ensuremath{\mathcal{SC}}}
\newcommand{\constraintArrival}{\ensuremath{\mathit{arrival}}}
\newcommand{\constraintDeparture}{\ensuremath{\mathit{departure}}}
\newcommand{\constraintVisit}{\ensuremath{\mathit{visit}}}
\newcommand{\constraintTransfer}{\ensuremath{\mathit{transfer}}}
\newcommand{\constraintWait}{\ensuremath{\mathit{wait}}}

\newcommand{\railwaySystem}{\ensuremath{\mathcal{S}}}

\newcommand{\imitator}{\textsf{IMITATOR}}

\newcommand{\eg}{e.g.,\xspace}

\newcommand{\ie}{i.e.,\xspace}
\newcommand{\suchthat}{s.t.\xspace}

\usepackage{orcidlink}
\renewcommand{\orcidID}[1]{\orcidlink{#1}}

\usepackage{fontawesome}
\newcommand{\homepage}[1]{\href{#1}{\color{gray}\faHome}}
\title{Tuning Trains Speed in Railway Scheduling\thanks{%
	\AuthorVersion{%
	This is the author version of the manuscript of the same name published in the proceedings of the 25th International Conference on Formal Engineering Methods ({ICFEM} 2024).
	The final version is available at \href{https://www.doi.org/10.1007/978-981-96-0617-7_3}{\nolinkurl{10.1007/978-981-96-0617-7_3}}.
	}%
	This work is partially supported by ANR BisoUS (ANR-22-CE48-0012).
	}
}
\titlerunning{Tuning Trains Speed in Railway Scheduling}
\author{\'Etienne Andr\'e\inst{1,2}%
	\homepage{https://lipn.univ-paris13.fr/~andre/}%
	\orcidID{0000-0001-8473-9555}
}
\institute{Université Sorbonne Paris Nord, LIPN, CNRS UMR 7030, \LongVersion{F-93430 }Villetaneuse, France
\and
Institut Universitaire de France (IUF)
}

\authorrunning{É.\ André}

\begin{document}
\sloppy

\maketitle{}
\begin{abstract}
	Railway scheduling consists in ensuring that a set of trains evolve in a shared rail network without collisions, while meeting schedule constraints.
	This problem is notoriously difficult, even more in the case of uncertain or even unknown train speeds.
	We propose here a modeling and verification approach for railway scheduling in the presence of uncertain speeds, encoded here as uncertain segment durations.
	We formalize the system and propose a formal translation to PTAs.
	As a proof of concept, we apply our approach to benchmarks, for which we synthesize using \imitator{} suitable valuations for the segment durations.

		\keywords{Railway scheduling \and Timed automata \and Parameter synthesis \and \imitator{}.}
\end{abstract}
\section{Introduction}\label{section:introduction}

Railway scheduling consists in ensuring that a set of trains evolve in a shared rail network without collisions, while meeting local or global, absolute or relative timing constraints.
This problem is notoriously difficult, and even more in the case of uncertain or even unknown train speeds, for which the solution needs to exhibit (or \emph{synthesize}) speeds for which the schedule constraints are met without collisions.
This becomes even more tricky when the schedule constraints (specifying, \eg{} the time difference between two events in the network) become themselves uncertain or unknown.

\paragraph{Contributions}
In this paper, we offer a modeling and verification framework for railway scheduling in the presence of uncertain speeds, modeled using uncertain segment durations.
Our railway model is close to that of~\cite{KR23} with some differences and simplifications: we consider a set of trains evolving in a shared network made of a double-vertex graph modeling segments and stations.
Segments have a length and a maximum speed (which can be refined using the maximum speed of trains); such lengths and speeds are here encoded using traversal durations.
Compared to~\cite{KR23}, we notably extend the model with the ability to express uncertain or unknown speeds (and therefore durations).
As target formalism for specification and verification, we choose parametric timed automata (PTAs)~\cite{AHV93}, an extension of timed automata (TAs)~\cite{AD94} with unknown timing constants, allowing to model variability and uncertainty.
Our contributions are three-fold:
\begin{enumerate}
	\item a formal modeling of the train trajectory problem under uncertain speeds;
	\item a translation scheme from our formal model into PTAs; and
	\item a set of experiments to show the applicability of our approach.
\end{enumerate}

\paragraph{Outline}
We review related works in \cref{section:related}.
We recall necessary preliminaries in \cref{section:preliminaries}.
We formalize our railway model (and the main problem) in \cref{section:model}.
Our translation to PTAs is described in \cref{section:translation}.
As a proof of concept, we apply our translation to benchmarks in \cref{section:experiments}.
We conclude in \cref{section:conclusion}.

\section{Related works}\label{section:related}

A number of works attempt to formalize railway scheduling problems using formal methods, with different model assumptions, and different target formalisms.
\LongVersion{%

}%
	In~\cite{NGPBVMGMV14,BDGMM17}, the focus is on the formalization of railway control systems using extensions of hierarchical state machines called ``Dynamic STate Machines'' (DSTMs).
In~\cite{VanitAnunchai10,VanitAnunchai18}, colored Petri nets are used to model railway interlocking tables, with applications to Thai railway stations.
\LongVersion{%

}%
Recent works such as~\cite{LCJS21,KR23} use SAT techniques, with \cite{KR23} modeling continuous dynamics in a quite involved way.

Timed automata are a particularly well-suited formalism to model such problems, due to their ability to model concurrent and timed behaviors.
Therefore, a number of works (such as \cite{YWK13,KASM16,ABRMA19,KLLS19,NTPAW19,LTH20}) are interested in scheduling or train interlocking problems.
Timing uncertainty is not considered though.

In~\cite{CWZT18}, so-called parametric timed automata (differing from usual PTAs~\cite{AHV93}, as events can be parametrized too) are used to build monitors with variability in order to perform runtime verification of computer-based interlocking systems; an application to Beijing metro line~7 is briefly studied.

In contrast to these works, we address here uncertain or unknown segment traversal durations; we allow in addition for parametric schedule constraints.

Beyond the specific application to railways, planning and scheduling using TAs was considered in, \eg{}~\cite{KMH01,AAM06,AM12}\LongVersion{, %
with extensions with prices/costs considered in \cite{BLR05,RLS06}}.
Scheduling in the presence of uncertainty was addressed in some works using parametric timed automata,
	including
		scheduling problems with applications to the aerospace~\cite{FLMS12,ACFJL21},
		or
		schedulability under uncertainty for uniprocessor environments~\cite{Andre17FMICS}.
\section{Preliminaries}\label{section:preliminaries}

We denote by $\setN, \setZ, \setQgeqzero, \setRgeqzero$ the sets of non-negative integers, integers, non-negative rationals and non-negative reals, respectively.
Let ${\compOp} \in \{<, \leq, =, \geq, >\}$.

\LongVersion{%
\subsection{Clocks, parameters and constraints}
}

\LongVersion{%
\subsubsection{Clocks}
}%
\emph{Clocks} are real-valued variables that all evolve over time at the same rate.
Throughout this paper, we assume a finite set~$\ClockSet = \{ \clocki{1}, \dots, \clocki{\ClockCard} \} $ of \emph{clocks}.
A \emph{clock valuation} is a function
$\clockval : \ClockSet \to \setRgeqzero$, assigning a non-negative value to each clock.
We write $\ClocksZero$ for the clock valuation assigning $0$ to all clocks.
Given a constant $d \in \setRgeqzero$, $\clockval + d$ denotes the valuation \suchthat\ $(\clockval + d)(\clock) = \clockval(\clock) + d$, for all $\clock \in \ClockSet$.

\LongVersion{%
\subsubsection{Parameters}
}%
A \emph{(timing) parameter} is an unknown rational-valued timing constant of a model.
Throughout this paper, we assume a finite set~$\ParamSet = \{ \parami{1}, \dots, \parami{\ParamCard} \} $ of \emph{parameters}.
A \emph{parameter valuation} $\pval$ is a function $\pval : \ParamSet \to \setQgeqzero$.

\LongVersion{%
\subsubsection{Constraint}
}%
A \emph{parametric clock constraint}~$\constraint$ is a conjunction of inequalities over $\ClockSet \cup \ParamSet$ of the form
$\clock \compOp \sum_{1 \leq i \leq |\ParamSet|} \alpha_i \parami{i} + d$, with
$\parami{i} \in \ParamSet$,
and
$\alpha_i, d \in \setZ$.
Given~$\constraint$, we write~$\clockval\models\pval(\constraint)$ if %
the expression obtained by replacing each~$\clock$ with~$\clockval(\clock)$ and each~$\param$ with~$\pval(\param)$ in~$\constraint$ evaluates to true.

\subsection{Parametric timed automata}
Parametric timed automata (PTAs) extend TAs with a finite set of timing parameters allowing to model unknown constants.

\begin{definition}[Parametric timed automaton~\cite{AHV93}]\label{definition:PTA}
	A PTA $\PTA$ is a tuple $\PTA = (\ActionSet, \LocSet, \locinit, \locfinal, \ClockSet, \ParamSet, \invariant, \EdgeSet)$, where:
	\begin{enumerate}%
		\item $\ActionSet$ is a finite set of actions;
		\item $\LocSet$ is a finite set of locations;
		\item $\locinit \in \LocSet$ is the initial location;
		\item $\locfinal \in \LocSet$ is the final location;
		\item $\ClockSet$ is a finite set of clocks;
		\item $\ParamSet$ is a finite set of parameters;
		\item $\invariant$ is the invariant, assigning to every $\loc\in \LocSet$ a parametric clock constraint $\invariant(\loc)$ (called \emph{invariant});
		\item $\EdgeSet$ is a finite set of edges  $\edge = (\loc,\guard,\action,\resets,\loc')$
		where~$\loc,\loc'\in \LocSet$ are the source and target locations, $\action \in \ActionSet$,
		$\resets\subseteq \ClockSet$ is a set of clocks to be reset, and $\guard$ is a parametric clock constraint (called \emph{guard}).
	\end{enumerate}%
\end{definition}

As often, we assume PTAs extended with discrete global variables such as integer- or Boolean-valued variables.
We also assume standard parallel composition of PTAs, synchronized on actions.
The parallel composition of $n$ PTAs is a~PTA.

\begin{definition}[Valuation of a PTA]\label{def:valuation-PTA}
	Given a parameter valuation~$\pval$, we denote by $\valuate{\PTA}{\pval}$ the non-parametric structure where all occurrences of a parameter~$\parami{i}$ have been replaced by~$\pval(\parami{i})$.
\end{definition}
\begin{remark}
	We have a direct correspondence between the valuation of a PTA and the definition of a TA.
	TAs were originally defined with integer constants in~\cite{AD94}, while our definition of PTAs allows \emph{rational}-valued constants.
	By assuming a rescaling of the constants (\ie{} by multiplying all constants in a TA by the least common multiple of their denominators), we obtain an equivalent (integer-valued) TA, as defined in~\cite{AD94}.
	So we assume in the following that $\valuate{\PTA}{\pval}$ is a~TA.
\end{remark}

\LongVersion{%
\paragraph{Concrete semantics of timed automata}

We recall the concrete semantics of a TA using a timed transition system (TTS).
}%
\begin{definition}[Semantics of a TA]
	Given a PTA $\PTA = (\ActionSet, \LocSet, \locinit, \locfinal, \ClockSet, \ParamSet, \invariant, \EdgeSet)$ and a parameter valuation~$\pval$
	the semantics of the TA $\valuate{\PTA}{\pval}$ is given by the TTS $\semantics{\valuate{\PTA}{\pval}} = (\StateSet, \concstateinit, \ActionSet \cup \setRgeqzero, \transition)$, with
	\begin{enumerate}
		\item $\StateSet = \big\{ (\loc, \clockval) \in \LocSet \times \setRgeqzero^\ClockCard \mid \clockval \models \invariant(\loc){\pval} \big\}$,
		\LongVersion{\item} $\concstateinit = (\locinit, \ClocksZero) $,
		\item  $\transition$ consists of the discrete and (continuous) delay transition relations:
		\begin{enumerate}
			\item discrete transitions: $(\loc,\clockval) \transitionWith{\edge} (\loc',\clockval')$,
			if $(\loc, \clockval) , (\loc',\clockval') \in \StateSet$, and there exists $\edge = (\loc,\guard,\action,\resets,\loc') \in \EdgeSet$, such that $\clockval'= \reset{\clockval}{\resets}$, and $\clockval\models\pval(\guard$).
			\item delay transitions: $(\loc,\clockval) \transitionWith{d} (\loc, \clockval+d)$, with $d \in \setRgeqzero$, if $\forall d' \in [0, d], (\loc, \clockval+d') \in \StateSet$.
		\end{enumerate}
	\end{enumerate}
\end{definition}

Moreover we write $(\loc, \clockval)\longuefleche{(d, \edge)} (\loc',\clockval')$ for a combination of a delay and discrete transition if
$\exists  \clockval'' :  (\loc,\clockval) \transitionWith{d} (\loc,\clockval'') \transitionWith{\edge} (\loc',\clockval')$.

Given a TA~$\TA$ with concrete semantics $\semantics{\TA}$, we refer to the states of~$\StateSet$ as the \emph{concrete states} of~$\TA$.
A \emph{run} of~$\TA$ is an alternating sequence of concrete states of~$\TA$ and pairs of edges and delays starting from the initial state $\concstateinit$ of the form
$(\loci{0}, \clockval_{0}), (d_0, \edge_0), (\loci{1}, \clockval_{1}), \cdots$
with
$i = 0, 1, \dots$, $\edge_i \in \EdgeSet$, $d_i \in \setRgeqzero$ and
$(\loci{i}, \clockval_{i}) \longuefleche{(d_i, \edge_i)} (\loci{i+1}, \clockval_{i+1})$.

\section{Railway system model}\label{section:model}

We formalize here our railway system model.
Our railway model is inspired by that of~\cite{KR23}, with some differences that will be highlighted.
A key difference is the ability of our model to define \emph{parametric} durations.
We also propose a more formal definition of the system.

\subsection{Rail network graph}

The infrastructure is modeled using a double-vertex graph~\cite{Montigel92}, with nodes and segments.
Nodes can be normal nodes (not allowing stopping) or stations (where trains may choose to stop or not).
Segments have a length and a speed limit, encoded here using a segment traversal time (which can be exceeded for slower trains that have a speed limit smaller than the segment maximal speed).
Boundary nodes are start or end nodes for the trains.
As in~\cite{KR23}, we do not model slope, angle or tunnels.
However, cycles can be encoded, as opposed to~\cite{LCJS21} where this is not immediate. %

We assume that segments are bidirectional, that at most one train is allowed in a segment, and that each segment is longer than any train; as a consequence, a train can occupy at most two segments at once.
As in~\cite{KR23},
``to support modeling of railway junctions, nodes of the graph have two sides (illustrated by black and blue or red colors in \cref{figure:example-rail-network-graph}).
In order to avoid physically impossible (\eg{} too sharp) turns, a train has to visit both sides when transferring via such a double-sided node.''
Different from~\cite{KR23}, we model segment length and speed using a \emph{traversal time}; similarly, since trains can occupy two segments at the same time, we model the time needed to completely move from one segment to the next one using another traversal time.
These times are minimum, as slower trains can potentially define longer times for each segment and pairs of segments (see \cref{definition:train}).
\LongVersion{%

}%
Formally:

\begin{definition}[rail network graph]
	A \emph{rail network graph} is a tuple $\railGraph = (\railNodes, \railBoundaries, \railStations, \railSegments, \railDuration, \railDurationPair, \railTransitions)$ where %
	\begin{itemize}
		\item $\railNodes$ is the set of nodes,
		\item $\railBoundaries \subseteq \railNodes$ is the set of boundary nodes,
		\item $\railStations \subseteq \railNodes$ is the set of stations,
		\item $\railSegments$ is the set of segments,
		\item $\railDuration : \railSegments \to \setQgeqzero \cup \ParamSet$ assigns a (possibly parametric) duration to each segment,
		\item $\railDurationPair : (\railSegments \times \railSegments) \topartial \setQgeqzero \cup \ParamSet$ assigns a (possibly parametric) duration to each pair of consecutive segments,
		and
		\item $\railTransitions \in 2^{\railSegments} \times \railNodes \times 2^{\railSegments}$ is the set of transitions.
	\end{itemize}
\end{definition}

Transitions encode the way trains can move via nodes.
For example, given a transition $(l, \railNode, r) \in \railTransitions$, a train can move from any segment $\railSegment \in l$ to any segment $\railSegment' \in r$ via $\railNode$ (or the opposite way).

\begin{example}
	Consider the rail network graph in \cref{figure:example-rail-network-graph}.
	The graph contains 4 boundary nodes (A, B, C, D) and 3 stations (depicted in red, and labeled with a number).
	Other nodes are normal nodes.
	Segments are labeled with a number identifying them.
	We can assume for example that the minimum time to traverse segment~1 is set to~8, the time to move from segment~1 to~2 is~2, while the time to move from segment~1 to~3 is~1 (values not depicted in \cref{figure:example-rail-network-graph}).
\end{example}
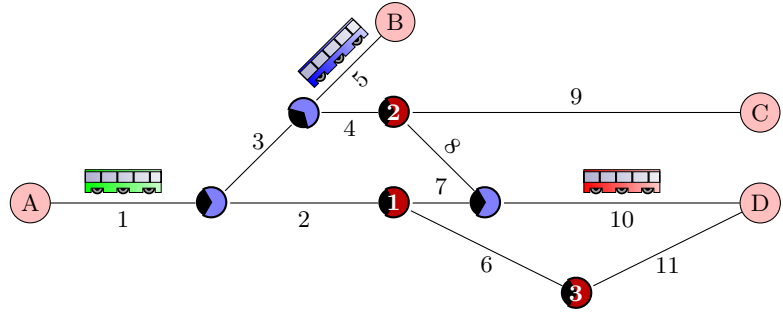
\begin{figure}[tb]

	\centering

	\begin{tikzpicture}[railway, scale=1.2, xscale=1, yscale=1]

		\node[boundary] at (0, 0) (A) {A};

		\node[railwaynode] at (2, 0) (N1) {\railwaynode{scale=.2,rotate=-180}};

		\node[railwaynode] at (3, 1) (N2) {\railwaynode{scale=.2,rotate=-135}};

		\node[station] at (4, 0) (TS1) {\station{scale=.2,rotate=-180}{1}};

		\node[station] at (4, 1) (TS2) {\station{scale=.2,rotate=-180}{2}};

		\node[railwaynode] at (5, 0) (N3) {\railwaynode{scale=.2,rotate=-180}};

		\node[station] at (6, -1) (TS3) {\station{scale=.2,rotate=-180}{3}};

		\node[boundary] at (4, 2) (B) {B};

		\node[boundary] at (8, 0) (D) {D};

		\node[boundary] at (8, 1) (C) {C};

		\path (A) edge node[] {\trainTikZ{.8}{green}} node[below]{1} (N1);
		\path (N1) edge node[above]{3} (N2);
		\path (N1) edge node[below]{2} (TS1);
		\path (N2) edge [sloped] node {\trainTikZ{.8}{blue}} node[below]{5} (B);
		\path (N2) edge node[below]{4} (TS2);
		\path (TS1) edge node[above]{7} (N3);
		\path (TS1) edge node[below]{6} (TS3);
		\path (TS2) edge[sloped] node[above]{8} (N3);
		\path (TS2) edge node[above]{9} (C);
		\path (N3) edge node[] {\trainTikZ{.8}{red}} node[below]{10} (D);
		\path (TS3) edge node[below]{11} (D);

	\end{tikzpicture}
	\caption{An example of a rail network graph with 3 trains \cite{KR23}}
	\label{figure:example-rail-network-graph}

\end{figure}
\subsection{Trains}
A train is characterized by its velocity limit and its connection.
Different from~\cite{KR23}, weight, acceleration and deceleration are not encoded; we assume they can be incorporated in segment durations.
Trains always drive at their maximum possible speed; they can however stop arbitrarily long in stations.

\subsubsection{Connection constraints}
As in~\cite{KR23}, a \emph{connection} is a mapping of a train to a non-empty list of nodes that must be visited in the given order.
Only the first and the last element can be boundary nodes.
The list of nodes must start with the boundary node denoting the starting point of the train.
The list may then contain nodes that must be visited; if a node is a station, then the train can stop at this station.
Trains can only stop at stations part of the connection.
If the last node of the list is a boundary node, then the train must end in this node.
If the last node of the connection is not a boundary node, then the train can end in any boundary node.

Each train has exactly one connection.

\begin{example}[train connections~\cite{KR23}]
	Consider the green train from \cref{figure:example-rail-network-graph}.
	Assume its connection is $[A, 3]$.
	This connection denotes that the green train must start at node~A, must stop at train station~3, but cannot stop at train station~1 because it is not part of the connection.
	The train can end in any boundary node (even though, considering the graph topology, only~D can be an end node considering the connection).

	Consider the red train from \cref{figure:example-rail-network-graph}.
	Assume its connection is $[D, A]$.
	This connection denotes that the train must depart from~D, and reach~A without stopping at any intermediate station; note that there are three paths allowing this connection.
\end{example}

\LongVersion{%
\subsubsection{Trains}
}%
\begin{definition}[train]\label{definition:train}
	Given a rail network graph $\railGraph = (\railNodes, \railBoundaries, \railStations, \railSegments, \railDuration, \railDurationPair, \railTransitions)$, %
	a \emph{train} over $\railGraph$ is a triple $\train = (\trainDuration, \trainDurationPair, \trainConnection)$ where %
	\begin{itemize}
		\item $\trainDuration : \railSegments \to \setQgeqzero \cup \ParamSet$  assigns a possibly parametric duration to each segment,
		\item $\trainDurationPair : (\railSegments \times \railSegments) \topartial \setQgeqzero \cup \ParamSet$  assigns a possibly parametric duration to each pair of consecutive segments,
		and
		\item $\trainConnection \in \railNodes^*$ is the train connection.
	\end{itemize}
\end{definition}

Given a segment, a train drives at its maximum speed depending on the network conditions: that is, the segment duration for this train is the \emph{maximum} between the segment duration specified by the network ($\railDuration$) and the segment duration specified by the train ($\trainDuration$)---and similarly for pairs of consecutive segments.

\subsection{Schedule constraints}

We formalize the schedule constraints from~\cite{KR23}, allowing to compare the time when a train arrives or departs from a node:
$\constraintArrival(\train, \railNode)$ (resp.\ $\constraintDeparture(\train, \railNode)$) denotes the time when train~$\train$ arrives at (resp.\ leaves) node~$\railNode$.
We use as generic notation $\constraintVisit(\train, \railNode)$ to denote arrival or departure.
We define three forms of constraints, detailed in the following.
An originality of our approach is that we also allow for \emph{parametric} constraints.

\subsubsection{Ordering constraints}
Ordering constraints constrain the order in which visits should be made.
They are of the form
\[ \constraintVisit_1(\train_1, \railNode_1) \compOp \constraintVisit_2(\train_2, \railNode_2) \text{.} \]

\subsubsection{Absolute timing constraints}
Absolute timing constraints constrain the visit of a node at an absolute time.
They are of the form
\[ \constraintVisit(\train, \railNode) \compOp d\text{, with }d \in \setQgeqzero \cup \ParamSet\text{.}\]

\subsubsection{Relative timing constraints}
Relative timing constraints constrain the time between two visits.
Let $\constraintTransfer\big(\constraintVisit_1(\train_1, \railNode_1), \constraintVisit_2(\train_2, \railNode_2)\big) \coloneq \constraintVisit_2(\train_2, \railNode_2) - \constraintVisit_1(\train_1, \railNode_1)$.
Then relative timing constraints are of the form
\[ \constraintTransfer\big(\constraintVisit_1(\train_1, \railNode_1), \constraintVisit_2(\train_2, \railNode_2)\big) \compOp d \text{, with }d \in \setQgeqzero \cup \ParamSet\text{.}\]

Finally, let us define $\constraintWait(\train, \railNode) \coloneq \constraintTransfer\big(\constraintArrival(\train, \railNode), \constraintDeparture(\train, \railNode)\big)$.

\begin{example}[schedule constraints]
	We formalize in the following some of the informal examples from~\cite{KR23}.
	The fact that the blue train must start before the green train can be encoded using
	$\constraintDeparture(\trainBlue, A) \leq \constraintDeparture(\trainGreen, A)$.
	\LongVersion{%

	}%
	The fact that the red train starts before the green train approaches node~1 can be encoded using
	$\constraintDeparture(\trainRed, D) \leq \constraintArrival(\trainGreen, 1)$.
	\LongVersion{%

	}%
	The fact that the red train must reach~A within 10~time units after entering the network can be encoded using
	$\constraintTransfer \big(\constraintDeparture(\trainRed, D) , \constraintArrival(\trainRed, A)\big) \leq 10$.
	\LongVersion{%

	}%
	The fact that the green train must wait at node~3 for at least $\param$~time units can be encoded using
	$\constraintWait(\trainGreen, 3) \geq \param$.

\end{example}
\subsection{Constrained railway system}
\begin{definition}[constrained railway system]
	A \emph{constrained railway system} is a tuple $\railwaySystem = (\railGraph, \Trains, \ScheduleConstraints)$ where
	\begin{itemize}
		\item $\railGraph$ is a rail network graph,
		\item $\Trains$ is a set of trains over $\railGraph$,
		and
		\item $\ScheduleConstraints$ is a set of schedule constraints.
	\end{itemize}
\end{definition}
\subsection{Objective}
\defProblem
	{Train trajectory problem under uncertain speeds}
	{a constrained railway system}
	{%
	Synthesize segment durations and schedule constraints parameters such that all train connections and schedule constraints are met.
}
\section{Translation to parametric timed automata}\label{section:translation}
\subsection{Overview of the translation}
Our translation is modular, in the sense that each train and each schedule constraint is translated into a different PTA.
The system is made of the parallel composition of these PTAs, synchronized using the actions modeling the arrival of a train into a node, and the departure of a train from a node.

\subsection{Railway model and trains}

Due to the \emph{concurrent} and \emph{real-time} nature of the system, a simple discrete graph with, \eg{} a list of trains currently at each node, is not a suitable approach.
Instead, we choose a fully distributed approach, where each train evolves in its own representation of the rail network graph, in a continuous manner.
That is, we define for each train~$k$ a PTA (with a single clock~$\clock^k$), with the set of locations being made of the segments and the node of the rail network graph.

The mutual exclusion in segments and nodes is ensured using global Boolean variables, carefully tested and updated when attempting to enter, and when exiting a segment or node.
More precisely, the occupancy of each segment~$\railSegment_i$ is encoded by a Boolean variable~$\segfreei{i}$ (denoting that the segment is free).

\begin{figure}[tb]
	{\centering
	\begin{tikzpicture}[pta, scale=1, yscale=1]
		\node[location, initial] at (0,0) (segm_i) {$\railSegment_i$};
		\node[location] at (0, -3) (segm_i_j) {$\railNode_{ij}$};
		\node[location] at (0, -6) (segm_j) {$\railSegment_j$};

		\node[invariant, right, xshift=.5em] at (segm_i.east) {$\styleclock{\clock^k} \leq \max\big(\railDuration(\railSegment_i) , \trainDuration(\railSegment_i)\big)$};

		\node[invariant, right, xshift=.5em] at (segm_i_j.east) {$\styleclock{\clock^k} \leq \max\big(\railDurationPair(\railSegment_i, \railSegment_j) , \trainDurationPair(\railSegment_i, \railSegment_j)\big)$};

		\node[invariant, right, xshift=.5em] at (segm_j.east) {$\styleclock{\clock^k} \leq \max\big(\railDuration(\railSegment_j) , \trainDuration(\railSegment_j)\big)$};

		\path[edge] (segm_i) edge[] node[align=center,right] {$\styledisc{\segfreei{j}} = \BTrue$
		\\
		$\land \styleclock{\clock^k} = \max\big(\railDuration(\railSegment_i) , \trainDuration(\railSegment_i)\big)$
			\\
			$\styleact{arrival_{ij}^k}$
			\\
			$\styleclock{\clock^k} \assign 0$
			\\
			$\styledisc{\segfreei{j}} \assign \BFalse$
		} (segm_i_j)
			;

		\path[edge] (segm_i_j) edge[] node[align=center,right] {$\styleclock{\clock^k} = \max\big(\railDurationPair(\railSegment_i, \railSegment_j), \trainDurationPair(\railSegment_i, \railSegment_j)\big)$
			\\
			$\styleact{departure_{ij}^k}$
			\\
			$\styleclock{\clock^k} \assign 0$
			\\
			$\styledisc{\segfreei{i}} \assign \BTrue$
		} (segm_j)
			;
		\end{tikzpicture}

	}
	\caption{Modeling two consecutive segments $\railSegment_i$ and $\railSegment_j$ via node $\railNode_{ij}$ for train~$\train_k$}
	\label{figure:segments}
\end{figure}
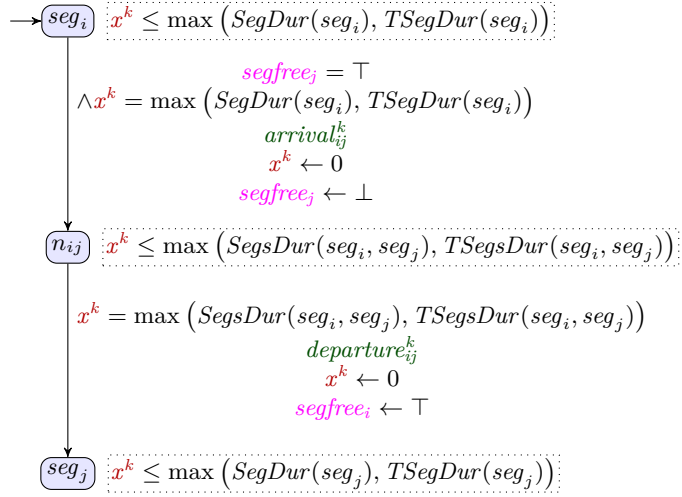

We give in \cref{figure:segments} the encoding of two consecutive segments $\railSegment_i$ and~$\railSegment_j$ via a node $\railNode_{ij}$ for a given train $\train_k = (\trainDuration, \trainDurationPair, \trainConnection)$.
As expected, the train can remain in a segment exactly $\max\big(\railDuration(\railSegment_i) , \trainDuration(\railSegment_i)\big)$ time units, and similarly in a location modeling the move of a segment to the next one (here location ``$\railNode_{ij}$'').
A train can move to the node between two segments only if the next segment ($\railSegment_j$) is free (``$\segfreei{j} = \BTrue$''),
	and the segment then becomes occupied (``$\segfreei{j} \assign \BFalse$'').
The actions ``$\textstyleact{arrival_{ij}^k}$'' and ``$\textstyleact{departure_{ij}^k}$'' are used to (potentially) synchronize with PTAs modeling schedule constraints.

If the node between $\railSegment_i$ and~$\railSegment_j$ is a station in which the train may stop (because it is part of its connection), then it is possible to stay longer in this node: in that case, the ``$=$'' sign in the guard between locations ``$\railNode_{ij}$'' and~``$\railSegment_j$'' in \cref{figure:segments} becomes~``$\geq$'', and the invariant of location~``$\railNode_{ij}$'' is removed.
\paragraph{Connections}
A connection is easily encoded using a discrete integer-valued variable:
a node occurring at the $n$th position of the connection can only be visited if $n-1$ nodes were visited by this train in the past, which is easily modeled by incrementing a local discrete integer.
The initial location of the train PTA is the node in which the train starts, and the final location is the node the train is supposed to reach at the end of its connection.

\subsection{Schedule constraints}
Ordering constraints are modeled exactly like connections, using discrete variables making sure the visits are performed in the specified order.

\begin{figure}[tb]

	\centering
	\begin{subfigure}[b]{0.5\textwidth}
		\centering
		\begin{tikzpicture}[pta, scale=1, yscale=1]
			\node[location, initial] at (0,0) (loc1) {$\loci{1}$};
			\node[location, final] at (2,0) (loc2) {$\loci{2}$};

			\path[edge] (loc1) edge[] node[above] {$\styleclock{\clockabs} \compOp d$} node[below]{$\styleact{visit_{n}^{t}}$} (loc2);

		\end{tikzpicture}

		\caption{Modeling constraint $\constraintVisit(\train, \railNode) \compOp d$}
		\label{figure:translation:visit-abs}
	\end{subfigure}
	\hfill
	\begin{subfigure}[b]{0.48\textwidth}
		\centering
		\begin{tikzpicture}[pta, scale=1, yscale=1]
			\node[location, initial] at (0,0) (loc1) {$\loci{1}$};
			\node[location] at (2,0) (loc2) {$\loci{2}$};
			\node[location, final] at (4,0) (loc3) {$\loci{3}$};

			\path[edge] (loc1) edge[] node[above] {$\styleact{visit_{n_1}^{t_1}}$} node[below]{$\styleclock{\clock} \assign 0$} (loc2);

			\path[edge] (loc2) edge[] node[above] {$\styleclock{\clock} \compOp d$} node[below]{$\styleact{visit_{n_2}^{t_2}}$} (loc3);
			\end{tikzpicture}

		\caption{Modeling constraint $\constraintTransfer\big(\constraintVisit_1(\train_1, \railNode_1), \constraintVisit_2(\train_2, \railNode_2)\big) \compOp d$}
		\label{figure:translation:transfer}
	\end{subfigure}

	\caption{Modeling schedule constraints}
\end{figure}
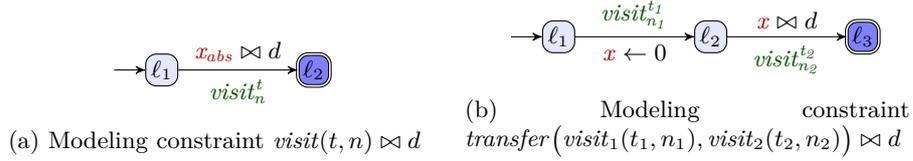

Each absolute timing constraint is modeled using a dedicated PTA, using a (single, global) clock $\clockabs$ measuring the absolute time, \ie{} never reset throughout the PTAs.
We give in \cref{figure:translation:visit-abs} the PTA modeling constraint $\constraintVisit(\train, \railNode) \compOp d$.
The PTA simply constrains action $\textstyleact{visit_{n}^{t}}$ to occur only whenever guard ``$\textstyleclock{\clockabs} \compOp d$'' is satisfied (recall that ``$\constraintVisit$'' stands for either ``$\constraintArrival$'' or ``$\constraintDeparture$'').

Each relative timing constraint is modeled using a dedicated PTA, using a local clock~$\clock$ measuring the relative time between different events.
We give in \cref{figure:translation:transfer} the PTA modeling the relative timing constraint $\constraintTransfer\big(\constraintVisit_1(\train_1, \railNode_1), \constraintVisit_2(\train_2, \railNode_2)\big) \compOp d$.
This PTA constrains the time difference between $\textstyleact{visit_{n_1}^{t_1}}$ and~$\textstyleact{visit_{n_2}^{t_2}}$ to be as specified by the constraint, using guard~``$\textstyleclock{\clock} \compOp d$''.

\subsection{Solving the train trajectory problem}

Given~$\PTA$ the PTA resulting of the parallel composition of the aforementioned PTAs, the trajectory problem without timing parameters is satisfied if the final location of all PTAs is reachable.
That is, each train reached its final destination, while all schedule constraints are satisfied.

In the presence of timing parameters, the set of parametric durations (modeling uncertain segment durations) for which all train connections and schedule constraints are met corresponds to the set of parameter valuations for which the final location of all PTAs is reachable.
This can be solved using reachability-synthesis, \ie{} the synthesis of timing parameters for which some PTA location is reachable~\cite{AHV93}.
While reachability-emptiness is in general undecidable~\cite{AHV93,Andre19STTT}, reachability-synthesis can be effectively solved under the assumption that there is no loop in the rail network graph.

\section{Experiments}\label{section:experiments}
As a proof of concept, we verify two constrained railway systems using the \imitator{} parametric timed model checker~\cite{Andre21}.
\imitator{} takes as input networks of parametric timed automata extended with a number of features, such as synchronization and discrete variables (used here).
Experiments were conducted on a Dell Precision 5570 with an Intel\textregistered{} Core\texttrademark{} i7-12700H with 32\,GiB memory running Linux Mint 21 Vanessa.
We used \imitator{} 3.4-alpha ``Cheese Durian'', build \texttt{feat/forall\_actions/788f551}.
Models (including durations for \cref{figure:example-rail-network-graph}) and results can be found at \url{https://www.imitator.fr/static/ICFEM24} and \url{https://doi.org/10.5281/zenodo.13789618}.

\begin{figure}[tb]

	\centering

	\begin{tikzpicture}[railway, xscale=1, yscale=1]

		\node[boundary] at (0, 0) (S) {S};

		\node[railwaynode] at (1, 0)  (S1)  {\railwaynode{scale=.2,rotate=-180}};
		\node[railwaynode] at (2, 1)  (N11) {\railwaynode{scale=.2,rotate=-135}};
		\node[railwaynode] at (2, 0)  (N21) {\railwaynode{scale=.2,rotate=-180}};
		\node[railwaynode] at (2, -1) (N31) {\railwaynode{scale=.2,rotate=-225}};
		\node[railwaynode] at (3, 0)  (E1)  {\railwaynode{scale=.2,rotate=0}};

		\node[railwaynode] at (4, 0)  (S2)  {\railwaynode{scale=.2,rotate=-180}};
		\node[railwaynode] at (5, 1)  (N12) {\railwaynode{scale=.2,rotate=-135}};
		\node[railwaynode] at (5, 0)  (N22) {\railwaynode{scale=.2,rotate=-180}};
		\node[railwaynode] at (5, -1) (N32) {\railwaynode{scale=.2,rotate=-225}};
		\node[railwaynode] at (6, 0)  (E2)  {\railwaynode{scale=.2,rotate=0}};

		\node[railwaynode] at (7, 0)  (S3)  {\railwaynode{scale=.2,rotate=-180}};
		\node[railwaynode] at (8, 1)  (N13) {\railwaynode{scale=.2,rotate=-135}};
		\node[railwaynode] at (8, 0)  (N23) {\railwaynode{scale=.2,rotate=-180}};
		\node[railwaynode] at (8, -1) (N33) {\railwaynode{scale=.2,rotate=-225}};
		\node[railwaynode] at (9, 0)  (E3)  {\railwaynode{scale=.2,rotate=0}};

		\node[boundary] at (10, 0) (E) {E};

		\path (S) edge (S1);

		\path (S1) edge (N11);
		\path (S1) edge (N21);
		\path (S1) edge (N31);
		\path (N11) edge (E1);
		\path (N21) edge (E1);
		\path (N31) edge (E1);

		\path (E1) edge (S2);

		\path (S2) edge (N12);
		\path (S2) edge (N22);
		\path (S2) edge (N32);
		\path (N12) edge (E2);
		\path (N22) edge (E2);
		\path (N32) edge (E2);

		\path (E2) edge (S3);

		\path (S3) edge (N13);
		\path (S3) edge (N23);
		\path (S3) edge (N33);
		\path (N13) edge (E3);
		\path (N23) edge (E3);
		\path (N33) edge (E3);

		\path (E3) edge (E);

		\end{tikzpicture}
	\caption{An example of a serial-parallel infrastructure, with $N_S = N_P = 3$ \cite{KR23}}
	\label{figure:example-serial-parallel}

\end{figure}
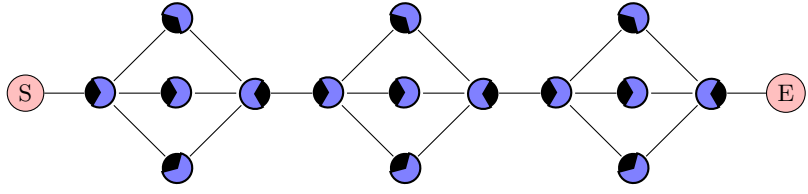

\paragraph{Running example}
We first consider the running example in \cref{figure:example-rail-network-graph}, without timing parameters.
\imitator{} derives in 1.24\,s that the train trajectory problem is satisfied, \ie{} there exists a schedule such that all trains meet their connections and schedule constraints.
Second, we add a parametric schedule constraint
\( \constraintVisit(\trainRed, A) \leq \paramRed \), where $\paramRed \in \ParamSet$.
That is, $\paramRed$ denotes the upper bound such that the red train reaches its destination.
\imitator{} derives in 1.91\,s the set of parameter valuations
\( \paramRed \geq 36 \), \ie{} the red train cannot be faster than 36~time units.
Third, we parametrize the minimum duration for segments 2 and~7, \ie{}
$\railDuration(\railSegment_2) = \parami{2}$
and
$\railDuration(\railSegment_7) = \parami{7}$,
with $\parami{2}, \parami{7} \in \ParamSet$.
\imitator{} derives in 9.83\,s the set of parameter valuations
\(
	(\paramRed \geq \parami{2} + 28)
	\lor
	(\paramRed \geq \parami{2} + \parami{7} + 20)
	\lor
	(\paramRed \geq 45)
\).
This gives the correct (sound and complete) condition over the segment durations and schedule constraints parameters such that all train connections and schedule constraints are met.
The fact that \imitator{} derives symbolic (continuous) sets of timing parameters is a major advantage over, \eg{} SMT solvers, that would typically derive (non-necessarily complete) discrete sets of specific valuations.
The symbolic and dense nature of our results comes from the underlying symbolic techniques for parameter synthesis using PTAs.

\paragraph{Scalability}
To evaluate the scalability of our approach, we consider a serial-parallel infrastructure, \ie{} a network with $N_S$ serially connected groups of $N_P$ identical parallel tracks with a station.
Connections only include $[S, E]$, \ie{} trains cannot stop at any station, and are free to use any path.
This variety of choices for each train obviously leads to an exponential blowup.
We consider various values for the number~$N_T$ of trains in the infrastructure, and for the number of groups~$N_S$ and parallel tracks~$N_P$;
as in~\cite{KR23}, we fix $N_S = N_P$.
An example with $N_S = N_P = 3$ is given in \cref{figure:example-serial-parallel}: $N_S = 3$ denotes the three groups from $S$ to~$E$ (from left to right), while $N_P = 3$ denotes the three options (drawn vertically) to traverse one group.

We reuse two scenarios from~\cite{KR23}: the ``nop'' scenario does not contain any parametric duration, nor any schedule constraint; however, contrarily to~\cite{KR23}, we add one parametric absolute timing constraint:
\(\constraintArrival(\train_{N_T}, E) \leq \paramJ\),
with $\paramJ \in \ParamSet$ (in~\cite{KR23}, $\paramJ$ is a constant manually tuned).
That is, we measure the end-to-end time from the first train leaving~$S$ to the last train reaching~$E$.

The ``last'' scenario in~\cite{KR23} additionally ensures that the last train takes less than~$\paramBound$ time units between its departure and arrival.
Again, we parametrize this value instead of manually enumerating it, by adding the following relative timing constraint:
\(\constraintTransfer\big(\constraintDeparture(\train_{N_T}, S), \constraintArrival(\train_{N_T}, E)\big) \leq \paramBound\),
with $\paramBound \in \ParamSet$.

Note that a direct comparison with the experiments of~\cite{KR23} would probably not make sense since
\begin{ienumerate}%
	\item the model is not the same (on the one hand, a more involved dynamics is considered in~\cite{KR23} and, on the other hand, we allow for more flexible durations and schedule constraints),
	\item the segment durations are not given in~\cite{KR23}
	and, most importantly,
	\item we \emph{synthesize} correct valuations while \cite{KR23} only \emph{verifies} the system for constant values.
\end{ienumerate}%

We give in \cref{table:XP:nop} the results for the ``nop'' scenario: we give from left to right the numbers of groups (and parallel tracks), of trains, of PTAs in the translated model, of clocks, of parameters, of generated states during the analysis;
we finally give the synthesized value for~$\paramJ$ and the computation time.
``\cellTO{}'' denotes timeout after 1\,800\,s.
Similarly, we give in \cref{table:XP:last} the results for the ``last'' scenario with, as additional column, the synthesized bound ``$\paramBound$'' for the relative timing constraint.

While the computation time is clearly exponential, which is not a surprise considering the way we designed this scalability test, a positive outcome is that we get interesting results for up to~4 trains or up to~4 groups of 4~parallel tracks, a rather elaborate situation---especially in a parametric setting with unknown timing bounds in the schedule constraints.

A difference with \cite{KR23,LCJS21} is that we can automatically synthesize the bound between the first train departure and the last train arrival, while they are manually iterated in~\cite{KR23,LCJS21}.
The second parameter (``$\paramBound$'') is simply tested in~\cite{KR23} against 3 values ($10$, $10^2$, $10^3$) without attempting to synthesize a tight valuation.

\begin{table}[tb]
\caption{Experiments}\label{table:XP}
	\setlength{\tabcolsep}{1pt} %
	\begin{subfigure}[b]{0.45\textwidth}
	\caption{Scenario ``nop''}\label{table:XP:nop}
	{\centering
		\begin{tabular}{| c | c | c | c | c | r | r | r |}
			\hline
			\rowHeader{}$N_S$ & $N_T$ & $|\PTA|$ & $|\ClockSet|$ & $|\ParamSet|$ & \cellCenter{$|S|$} & \cellCenter{$\paramJ$} & \cellCenter{$t(s)$} \\
			\hline
			2 & 1 & 2 & 2 & 1 & 25    & 63 & 0.01  \\ %
			2 & 2 & 3 & 3 & 1 & 238   & 75 & 0.08  \\ %
			2 & 3 & 4 & 4 & 1 & 2323  & 87 & 2.68  \\
			2 & 4 & 5 & 5 & 1 & 22450 & 99 & 96.02 \\
			\hline
			3 & 1 & 2 & 2 & 1 & 51    & 92  & 0.01   \\ %
			3 & 2 & 3 & 3 & 1 & 1237  & 104 & 0.68   \\ %
			3 & 3 & 4 & 4 & 1 & 27385 & 116 & 137.61 \\
			3 & 4 & - & - & - & - & - & \cellTO{} \\
			\hline
			4 & 1 & 2 & 2 & 1 & 87    & 121 & 0.03 \\ %
			4 & 2 & 3 & 3 & 1 & 3195  & 133 & 3.28 \\
			4 & 3 & - & - & - & - & - & \cellTO{} \\
			\hline
		\end{tabular}

	}
	\end{subfigure}
	\hfill
	\begin{subfigure}[b]{0.5\textwidth}
	\caption{Scenario ``last''}\label{table:XP:last}
	{\centering
		\begin{tabular}{| c | c | c | c | c | r | r | r | r |}
			\hline
			\rowHeader{}$N_S$ & $N_T$ & $|\PTA|$ & $|\ClockSet|$ & $|\ParamSet|$ & \cellCenter{$|S|$} & \cellCenter{$\paramJ$} & \cellCenter{$\paramBound$} & \cellCenter{$t(s)$} \\
			\hline
			2 & 1 & 2 & 3 & 2 & 25    & 63 & 63 & 0.01   \\ %
			2 & 2 & 3 & 4 & 2 & 238   & 75 & 63 & 0.12   \\ %
			2 & 3 & 4 & 5 & 2 & 2323  & 87 & 63 & 3.40   \\
			2 & 4 & 5 & 6 & 2 & 22450 & 99 & 63 & 113.41 \\
			\hline
			3 & 1 & 2 & 3 & 2 & 51    & 92  & 92 & 0.02   \\ %
			3 & 2 & 3 & 4 & 2 & 1052  & 104 & 92 & 0.91   \\
			3 & 3 & 4 & 5 & 2 & 27385 & 116 & 92 & 157.78 \\
			3 & 4 & - & - & - & - & - & - & \cellTO{} \\
			\hline
			4 & 1 & 2 & 3 & 2 & 87   & 121 & 121 & 0.04 \\ %
			4 & 2 & 3 & 4 & 2 & 3195 & 133 & 121 & 4.64 \\
			4 & 3 & - & - & - & - & - & - & \cellTO{} \\
			\hline
		\end{tabular}

	}
	\end{subfigure}
\end{table}
\section{Conclusion and perspectives}\label{section:conclusion}

We presented a formal model for verifying constrained railway systems in the presence of unknown durations, not only to model segment traversal times, but also to be used in relative and absolute schedule constraints.
Our translation to PTAs allowed us to verify benchmarks using \imitator{}, and to derive internal segment durations and optimal values for schedule constraints.

We believe our framework, although simple, can serve as a preliminary basis for more involved settings.
Notably, modeling acceleration and deceleration would be an interesting enhancement, possibly using piecewise discretization to keep the linear nature of our framework.
Taking energy consumption into consideration would be another interesting future work, \eg{} with an optimality criterion, perhaps integrating our setting with other approaches such as~\cite{BBFLMR21,LRS21}.
In addition, tackling the exponential blowup could be partially achieved using partial order or symmetry reductions, since these models are heavily symmetric.
Finally, we used here an \emph{ad~hoc} modeling language; integrating this framework into standard domain-specific languages will be an interesting extension.

\LongVersion{\section*{Acknowledgments}}%
\ShortVersion{\paragraph{Acknowledgments}}%
I would like to thank Tomáš Kolárik and Stefan Ratschan for introducing me to their work,
and an anonymous reviewer for helpful comments.
The colored trains drawn using \LaTeX{} TikZ in \cref{figure:example-rail-network-graph} are designed by \href{https://tex.stackexchange.com/users/39222/cfr}{cfr} from \href{https://tex.stackexchange.com/questions/235535/draw-a-car-profile-with-tikz}{stackexchange}.

\ifdefined\VersionWithComments
	\newcommand{\CCIS}{Communications in Computer and Information Science}
	\newcommand{\ENTCS}{Electronic Notes in Theoretical Computer Science}
	\newcommand{\FAC}{Formal Aspects of Computing}
	\newcommand{\FundInf}{Fundamenta Informaticae}
	\newcommand{\FMSD}{Formal Methods in System Design}
	\newcommand{\IJFCS}{International Journal of Foundations of Computer Science}
	\newcommand{\IJSSE}{International Journal of Secure Software Engineering}
	\newcommand{\IPL}{Information Processing Letters}
	\newcommand{\JAIR}{Journal of Artificial Intelligence Research}
	\newcommand{\JLAP}{Journal of Logic and Algebraic Programming}
	\newcommand{\JLAMP}{Journal of Logical and Algebraic Methods in Programming} %
	\newcommand{\JLC}{Journal of Logic and Computation}
	\newcommand{\LMCS}{Logical Methods in Computer Science}
	\newcommand{\LNCS}{Lecture Notes in Computer Science}
	\newcommand{\RESS}{Reliability Engineering \& System Safety}
	\newcommand{\RTS}{Real-Time Systems}
	\newcommand{\SCP}{Science of Computer Programming}
	\newcommand{\SOSYM}{Software and Systems Modeling ({SoSyM})}
	\newcommand{\STTT}{International Journal on Software Tools for Technology Transfer}
	\newcommand{\TCS}{Theoretical Computer Science}
	\newcommand{\TOPLAS}{{ACM} Transactions on Programming Languages and Systems ({ToPLAS})}
	\newcommand{\ToPNoC}{Transactions on {P}etri Nets and Other Models of Concurrency}
	\newcommand{\TOSEM}{{ACM} Transactions on Software Engineering and Methodology ({ToSEM})}
	\newcommand{\TSE}{{IEEE} Transactions on Software Engineering}
\else
	\newcommand{\CCIS}{CCIS}
	\newcommand{\ENTCS}{ENTCS}
	\newcommand{\FAC}{FAC}
	\newcommand{\FundInf}{FI}
	\newcommand{\FMSD}{FMSD}
	\newcommand{\IJFCS}{IJFCS}
	\newcommand{\IJSSE}{IJSSE}
	\newcommand{\IPL}{IPL}
	\newcommand{\JAIR}{JAIR}
	\newcommand{\JLAP}{JLAP}
	\newcommand{\JLAMP}{JLAMP}
	\newcommand{\JLC}{JLC}
	\newcommand{\LMCS}{LMCS}
	\newcommand{\LNCS}{LNCS}
	\newcommand{\RESS}{RESS}
	\newcommand{\RTS}{RTS}
	\newcommand{\SCP}{SCP}
	\newcommand{\SOSYM}{{SoSyM}}
	\newcommand{\STTT}{STTT}
	\newcommand{\TCS}{TCS}
	\newcommand{\TOPLAS}{ToPLAS}
	\newcommand{\ToPNoC}{ToPNOC}
	\newcommand{\TOSEM}{ToSEM}
	\newcommand{\TSE}{TSE}
\fi

\ifdefined\VersionAuthor
	\renewcommand*{\bibfont}{\small}
	\printbibliography[title={References}]
\else
	\bibliographystyle{splncs04} %
	\bibliography{trains}
\fi

\end{document}